\documentclass[journal]{vgtc}                
\usepackage[dvipsnames]{xcolor}
\usepackage{amsmath,amsfonts,amssymb}
\usepackage[english]{babel}    
\usepackage{url}
\usepackage{chngcntr}          
\usepackage{listings}
\usepackage{comment}
\usepackage{gensymb}  
\usepackage{booktabs} 
\usepackage{siunitx} 

\usepackage{enumitem}
\usepackage[
singlelinecheck=false 
]{caption} 

\ieeedoi{10.1109/TVCG.2019.2934536}

\addto{\captionsenglish}{}

\DeclareMathAlphabet{\mathsfit}{T1}{\sfdefault}{\mddefault}{\sldefault} 

\ifpdf
  \pdfoutput=1\relax                   
  \usepackage{graphicx}                
  \DeclareGraphicsExtensions{.pdf,.png,.jpg,.jpeg} 
\else
  \usepackage{graphicx}                
  \DeclareGraphicsExtensions{.eps}     
\fi%

\graphicspath{{figures/}{pictures/}{images/}{./}} 

\usepackage{microtype}                 
\PassOptionsToPackage{warn}{textcomp}  
\usepackage{textcomp}                  
\usepackage{mathptmx}                  
\usepackage{times}                     
\usepackage{cite}                      
\usepackage{tabu}                      
\usepackage{booktabs}                  

\onlineid{1119}

\vgtccategory{Research}
\vgtcpapertype{Evaluation}

\title{Context Matters: A Theory of Semantic Discriminability for Perceptual Encoding Systems}

\author{Kushin Mukherjee, Brian Yin, Brianne E. Sherman, Laurent Lessard, and Karen B. Schloss}

\authorfooter{

\item  Kushin Mukherjee, Psychology and Wisconsin Institute for Discovery, University of Wisconsin--Madison . Email: kmukherjee2@wisc.edu.
\item  Brian Yin, Cognitive Science, University of California, Berkeley. Email:brianyin@berkeley.edu.
\item Brianne E. Sherman, Neurobiology and Wisconsin Institute for Discovery, University of Wisconsin--Madison,  Email: besherman2@wisc.edu.
\item  Laurent Lessard, Mechanical and Industrial Engineering, Northeastern University. Email: l.lessard@northeastern.edu.
\item  Karen B. Schloss, Psychology and Wisconsin Institute for Discovery, University of Wisconsin--Madison. Email: kschloss@wisc.edu.

}

\shortauthortitle{Mukherjee \MakeLowercase{\textit{et~al.}}: Semantic Discriminability for Visual Communication}

\abstract{
People's associations between colors and concepts influence their ability to interpret the meanings of colors in information visualizations. Previous work has suggested such effects are limited to concepts that have strong, specific associations with colors. However, although a concept may not be strongly associated with any colors, its mapping can be disambiguated in the context of other concepts in an encoding system. We articulate this view in semantic discriminability theory, a general framework for understanding conditions determining when people can infer meaning from perceptual features. Semantic discriminability is the degree to which observers can infer a unique mapping between visual features and concepts. Semantic discriminability theory posits that the capacity for semantic discriminability for a set of concepts is constrained by the difference between the feature-concept association distributions across the concepts in the set. We define formal properties of this theory and test its implications in two experiments. The results show that the capacity to produce semantically discriminable colors for sets of concepts was indeed constrained by the statistical distance between color-concept association distributions (Experiment 1). Moreover, people could interpret meanings of colors in bar graphs insofar as the colors were semantically discriminable, even for concepts previously considered ``non-colorable'' (Experiment 2). The results suggest that colors are more robust for visual communication than previously thought.
}


 \teaser{
   \centering
   \includegraphics[width=0.95\linewidth]{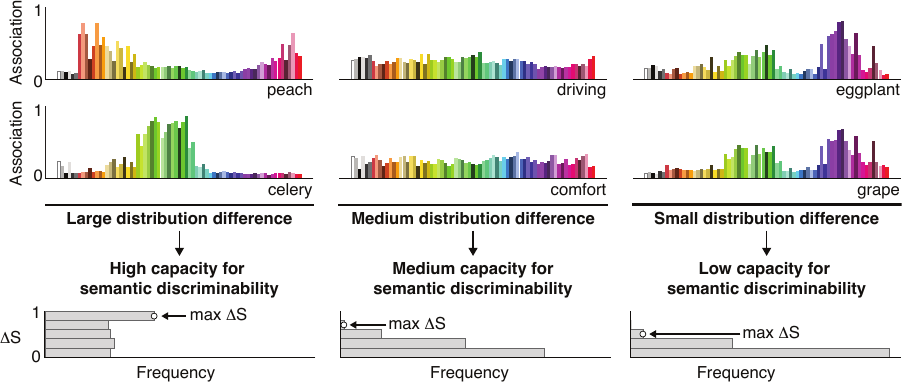}
   \caption{Color-concept association distributions for concept pairs with large, medium, and small distribution differences, resulting in high, medium, and low capacities for semantic discriminability, respectively (terms defined in Section \ref{sec:theory}). Color-concept association ratings were collected in Experiment~1 for the UW-71 colors (colored stripes in the plots, sorted by CIE LCh hue angle). 
   }\label{fig:assoc}
 }

 \nocopyrightspace

\begin{document}


\firstsection{Introduction} \label{sec:intro}
\maketitle
Bananas are shades of yellow, blueberries are shades of blue, and cantaloupes are shades of orange. It is well-established that color semantics influences people's ability to interpret information visualizations when those visualizations represent concepts that have specific, strongly associated colors (e.g., fruits). Such visualizations are easier to interpret if concepts are encoded with strongly associated colors (e.g., bananas encoded with yellow, not blue) \cite{lin2013,schloss2021}. But, how often do real-world visualizations really depict information about fruit, or other concepts with specific, strongly associated colors? If color semantics mainly influences interpretability for visualizations of concepts with specific, strongly associated colors (as previously suggested \cite{lin2013, setlur2016}), then scenarios in which color semantics matters would be severely limited.
\vspace{-4mm}

The present study suggests people's ability to infer meaning from colors is more robust than previously thought. Conditions arise in which people can interpret meanings of colors for concepts previously considered ``non-colorable''. Specifically this when the colors are semantically discriminable. Semantic discriminability for colors is the ability to infer unique mappings between colors and concepts based on colors and concepts alone (i.e., without using a legend)\cite{schloss2021}. This is distinct from \textit{semantic interpretability}, which is the ability to interpret the \textit{correct} mapping between colors and concepts, as specified in an encoding system (for further discussion of this distinction, see \cite{schloss2021} and Supplementary Material Section \ref{sec:SC_supp} in the present paper). The key question is, what determines whether it is possible to select semantically discriminable colors for a set of concepts? 

We address this question in \textit{semantic discriminability theory}, a new theory on constraints for generating semantically discriminable perceptual features for encoding systems that map perceptual features to concepts. We tested two hypotheses that arise from the theory. First, the capacity to create semantically discriminable color palettes for a set of concepts depends on the difference in color-concept association distributions \textit{between} those concepts, independent of properties of the concepts in isolation (Experiment 1). Second, people can accurately interpret mappings between colors and concepts for concepts previously considered “non-colorable,” to the extent that the colors are semantically discriminable (Experiment 2). We focus on color in this study, but present the theory in terms of perceptual features more generally because of its potential to extend to other types of visual features (e.g., shape, orientation, visual texture) and features in other perceptual modalities (e.g., sound, odor, touch). 

\textbf{Contributions.} This paper makes the following contributions: (1)~We define semantic discriminability theory (Section \ref{sec:theory}) and test hypotheses motivated by the theory in Experiments 1 and 2 (Sections \ref{sec:exp1}-\ref{sec:exp2}), and (2) We define a new metric for operationalizing distribution difference between sets of more than two concepts (Section \ref{sec:semdistthry}) and show that it predicts capacity for semantic discriminability (Section \ref{sec:exp1}).
\vspace{-2mm}

\section{Background} \label{sec:background}

Color is a strong cue for signaling meaning in nature and some argue that color vision evolved for the purpose of visual communication \cite{humphrey1976, hasantash2019, changizi2006, conway2009, thorstenson2018}. Historically, discussions on the role of color semantics in information visualization have tended to focus on few cases of typical associations (e.g., red for hot, green for grass) \cite{shah2002, brockmann1991, robinson:1952}. More recent work has sought to understand the potential and limitations of using color to communicate meaning in visualizations \cite{lin2013, setlur2016, schloss2018, schloss2021, bartram2017, anderson2021}. The semantics of color in visualizations operates on two main levels: meaning of a color palette as a whole \cite{bartram2017, jahanian2017, anderson2021} and meaning of the individual colors in a palette\cite{lin2013, setlur2016, schloss2018, schloss2021}. We focus on meanings of individual colors because that is central to the present work. People have expectations about how colors will map onto concepts, and visualizations that violate those expectations are harder to interpret, even if there is a legend \cite{lin2013, schloss2019, sibrel2020}. Thus, understanding these expectations is important for optimizing palette design for visual communication. 

\subsection{Color-concept associations}
Color-concept associations represent the degree to which individual colors are associated with individual concepts. 
Color-concept associations can be quantified using various methods, including human judgments \cite{ou2004, jonauskaite2019sun, jonauskaite2019machine, tham2020systematic, rathore2020, schloss2021, schloss2018}, image statistics \cite{lindner2012a, lindner2012b, lin2013, setlur2016, rathore2020}, and natural language corpora \cite{setlur2016, havasi2010}. Some approaches focused on identifying the strongest, or strongest few colors associated with a concept \cite{havasi2010, setlur2016, fang2017}, but color-concept associations can be treated as a continuous property over all possible colors in a perceptual color space \cite{schlossPICS, lindner2012a, lindner2012b, lin2013, rathore2020}. When quantifying color-concept associations over all of color space, researchers typically bin or sub-sample parts of the space to make measurements computationally tractable. An assumption is that the space is continuous, so nearby colors will have similar associations. Figure \ref{fig:assoc} shows examples of color-concept associations for colors systematically sampled over CIELAB space (see Experiment 1), plotted over one dimension (sorted by hue angle and chroma with achromatics at the beginning of the list). Perceptual color spaces are three-dimensional so this representation does not necessarily position perceptually similar colors in close proximity \cite{wyszecki1982}, but it does highlight how some concepts, like peach and celery, have specific, strongly associated colors, whereas other concepts, like driving and comfort, are more uniform (Figure \ref{fig:assoc}). We refer to this `peakiness' property as \textit{specificity} of the color-concept association distributions.\footnote{Specificity is similar to color diagnosticity \cite{tanaka1999}, but color diagnosticity concerns whether a concept has a single strongly associated color \cite{tanaka1999}, and specificity concerns the degree to which a concept is associated with some colors more than others in a color-concept association distribution.}

Questions remain concerning how color concept-associations are formed, but many have suggested that they are learned through experiences \cite{elliot2007color, witzel2011, rathore2020, tham2020systematic, jonauskaite2019sun} and may be continually updated through each new experience in the world \cite{schlossPICS}. Some color-concept associations are shared cross-culturally, and others are subject to cultural differences \cite{tham2020systematic, jonauskaite2019sun, jonauskaite2019machine}. We will consider the role of cultural differences with respect to the present work in the General Discussion. 

Color-concept associations contribute to people's expectations about the meanings of colors in information visualizations \cite{lin2013, schloss2018, schloss2021}, called \textit{inferred mappings}. However, associations and inferred mappings are not the same, and sometimes they conflict \cite{schloss2018}. We explain this point in Section \ref{sec:assignment} on assignment inference. 

\subsection{Colorabilty scores}

Some have suggested that the effectiveness of colors for encoding meaning is limited to concepts that have strong associations with particular colors \cite{setlur2016, lin2013, kay2001linguistics}. This idea is explained by invoking \textit{colorability} scores, which broadly measure how strongly \textit{individual} concepts can be mapped to specific colors. Generally, concepts with specific, strongly associated colors (`banana') are thought to be colorable, whereas more abstract concepts, such as `comfort' or `leisure', that lack such strongly associated colors, have been called non-colorable. 

Different methods have been used to define colorability. Lin et al. \cite{lin2013} quantified colorability by having participants assign colors to concepts and rate the strength of the assignment. The mean of these ratings over all colors for a concept was used to generate a colorability score for that concept. They found that participants were better at interpreting bar charts when palettes were optimized for color semantics compared to when palettes had the default Tableau color ordering, but this benefit was mostly limited to highly colorable concepts. Setlur and Stone \cite{setlur2016} quantified colorability with an automated method, using Google N-grams to determine how frequently a concept word co-occurred with basic color terms~\cite{berlin1969} in linguistic corpora. They then excluded concepts they found to be non-colorable when developing methods to optimize palette design.

These prior studies highlighted the importance of considering color semantics in palette design. However, our work suggests that restricting notions of colorability to concepts in isolation may have led to underestimating people's ability to infer meaning from colors in visualizations.

\subsection{Assignment inference}\label{sec:assignment}

\begin{figure}[tb]
 \centering
 \includegraphics[width=0.95\columnwidth]{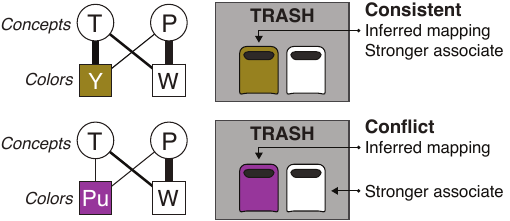}
 \caption{Distinction between color-concept associations and inferred mappings (figure based on \cite{schloss2018}). Left: Bipartite graphs show color-concept association strengths for concepts trash (T) and paper (P) with colors dark yellow (Y), white (W), and purple (Pu) (thicker edges connecting concepts and colors indicate stronger associations). Right: example trials where participants infer which color maps to trash.}
 \label{fig:rec}
 \vspace{-6mm}
\end{figure}

Evidence suggests that people's inferences about the  meanings of colors in encoding systems of visualizations do not merely depend on color-concept associations in isolation. We illustrate this point with an example from Schloss et al. \cite{schloss2018}. Participants saw pairs of unlabeled bins and were asked to choose which bin was for the target concept written at the top of the screen. Figure \ref{fig:rec} shows two examples when trash was the target concept. The other concept, not pictured here but judged on other trials, was paper. To the left of the example trials are bipartite graphs, which use line thickness to represent the association strength between each concept (trash, T, and paper, P) and each color in the corresponding trial. An easy way to approach this task would be to choose the color that is most strongly associated with trash within each trial (local assignment). Alternatively, participants could choose the color that results in maximizing association strengths of all color-concept pairings across trials (global assignment). 

In the top row of Figure \ref{fig:rec}, these two approaches lead to the same outcome. Locally, trash is more strongly associated with dark yellow (Y) than white (W). Globally, the assignment trash-yellow/paper-white has a larger overall association strength than trash-white/paper-yellow. Not surprisingly, participants inferred trash is mapped to dark yellow. However, in the bottom row, the two approaches lead to opposite outcomes. Locally, trash is more associated with white than purple (Pu), but globally the assignment trash-purple/paper-white has a larger overall association strength (greater total thickness of edges) than trash-white/paper-purple. Participants inferred that trash maps to purple, even though white was a more strongly associated alternative. Each trial was independent, so participants need not account for paper on trials for trash, but they did so nonetheless. This example highlights the important distinction between color-concept associations for a single color and concept, and inferred mappings between a color and concept in the context of an encoding system. 

Schloss et al. \cite{schloss2018} called this process of inferring mappings between colors and concepts \textit{assignment inference} because it is analogous to an \textit{assignment problem} in optimization. In assignment problems, every possible pairing of items in one category (e.g., colors) $i$ and another category (e.g., concepts) $j$ is given a numerical \emph{merit score} $m_{ij}$. Here, let's assume that larger scores indicate a more desirable pairing, but that is not always true (e.g., to optimize delivery route efficiency, merit might be delivery time and smaller scores would be better). Solving an assignment problem means finding the pairing of items
 that maximizes (or minimizes) the sum of the merit scores of all chosen pairs \cite{kuhn1955hungarian,munkres1957algorithms,burkard2012assignment}. 

Although assignment inference is analogous to assignment problems, they are not the same. Assignment problems have deterministic results, whereas assignment inference is stochastic---inferred mappings can vary among individuals and even within individuals over time. This stochasticity can be explained in terms of noise in people's color-concept associations affecting the outcome of assignments in assignment inference, depending on whether assignments are robust or fragile \cite{schloss2021}. In robust assignments, adding noise to the system (e.g., perturbing the color-concept association strengths) has no effect on the outcome, but in fragile assignments adding noise can change the outcome of the assignment. 

The robustness of an assignment in assignment inference can be understood as \emph{semantic discriminability}---the ability for people to infer a unique mapping between colors and concepts~\cite{schloss2021}. Evidence suggests that semantic discriminability predicts people's ability to interpret colors in encoding systems, independent of that predicted by perceptual discriminability and color-concept associations in isolation \cite{schloss2021}. We describe ways of operationalizing semantic discriminability in Section~\ref{sec:semdistthry} as they pertain to the present study.

So far, we focused on encoding systems with two concepts and colors, and implied that merit $m_{ij}$ in assignment inference is color-concept association strength (Figure \ref{fig:rec}). However, there are other possible ways to define merit, especially when there are more than two colors and concepts, as in the present study. Schloss et al. \cite{schloss2018} sought to understand which merit people use in assignment inference to study (1) how humans infer meaning from colors, and (2) how to design palettes that match people's expectations, making palettes more interpretable. To approach this goal, they created two definitions of merit. The \textit{isolated merit function} simply uses association strengths between items $i$ and $j$, $m_{ij} := a_{ij}$. The \textit{balanced merit function} is defined as
\begin{equation}\label{eq:balanced_merit}
 \vspace{-1mm}
m_{ij} := a_{ij} - \max_{k\neq j}a_{ik}.
 \vspace{-1mm}
\end{equation}
The balanced merit score for a given color-concept pair is the association strength for that pair, minus the association strength between that color and the next most strongly associated concept. In order for $m_{ij}$ to be large, color $i$ should be strongly associated with concept $j$ and weakly associated with all other concepts. (Note: in the case of two concepts and colors these two definitions reduce to the same outcome.)

Next, they generated color palettes using an assignment problem under each definition, with human color-concept association ratings as the input. Finally, they presented different participants with those palettes in the form of six unlabeled colored bins. Participants inferred which bin was for each of six objects: paper, plastic, trash, metal, compost, and glass. Responses were scored as ``correct'' interpretations if they matched the encoded mapping. Encoded mappings can be produced in different ways, including by designers, software defaults, or optimization algorithms \cite{lin2013, schloss2018, schloss2021}. Here, they were determined by the optimal assignments in assignment problems used to generate the palettes. The logic was that participants would be better at interpreting palettes generated using a merit function that more closely matched merit in assignment inference. Performance was better for the palette generated using the balanced merit function, which suggests that this was the function that better captured merit in assignment inference. Thus, we use balanced merit in the present study. 

Balanced merit can lead to unexpected assignments. For example, the bin for plastic was assigned a red color, even though red was weakly associated with plastic, because that color was more associated with plastic than with any of the other concepts. Thus, the assignment of plastic--red was interpretable. Given that weakly associated colors can prove useful when designing encoding systems, approaches that focus only on the top associates may be limited \cite{fang2017}. It is important to quantify associations between concepts and a large range of colors, not just the top few associates, when optimizing palette design \cite{rathore2020}.

\section{Semantic Discriminability Theory}\label{sec:theory}

Semantic discriminability theory characterizes the ability to generate semantically discriminable perceptual features for encoding a set of concepts. We begin with some key definitions.

    \textbf{Concept set:} This is the set of all concepts that are represented in an encoding system. These concepts could refer to any information that is categorical (e.g., food, weather, activities, places, and animals). We label concepts in the concept set using the index $j\in\{1,2,\dots,n\}$.
    
    \textbf{Feature source:} This is the set of all possible instances of a feature type. Perceptual color spaces (e.g., CIELAB) are well-defined feature sources for color, as they represent all colors humans can perceive \cite{wyszecki1982}.
    
    \textbf{Feature library:} This is a subset of candidate features from the feature source used in an encoding system. For example, the Tableau 20 colors or UW-58 colors \cite{rathore2020} are feature libraries if design is constrained to those groups of colors. We focus on a feature library defined over color, but they can be defined over any type of perceptual feature (e.g., shapes, sizes, textures). We label features in the feature library using the index $i\in \{1,2,\dots,N\}$.

    \textbf{Feature set:} This is a subset of features from the feature library, selected to encode a concept set. Feature sets can be constructed from any type of perceptual features (e.g., colors, shapes, sizes) \cite{bertin1983}. For colors, they are called ``palettes.'' If there are $n$ concepts, then the feature set should contain $n$ features.
    
    
     \subsection{Feature-concept association distributions}\label{sec:dists}
     
     Feature-concept association distributions represent the degree to which a given concept is associated with each feature in a feature library (see Figure \ref{fig:all_concepts}A. in the Supplementary Material). For color, these are color-concept association distributions. Feature-concept association distributions can be described as raw association values over the feature library (e.g., mean ratings, pixel counts, word counts). In this case, we write $a_{ij}$ to denote the association between feature $i\in\{1,\dots,N\}$ and concept $j\in\{1,\dots,n\}$. For each concept $j$, we also define \emph{normalized} associations $p_j(\cdot)$ as
     \begin{equation}\label{eq:normalized_pij}
     p_j(i) := \frac{a_{ij}}{\sum_{k=1}^N a_{kj}}
     \qquad\text{for: }i\in\{1,\dots,N\}.
     \end{equation}
    The list $\begin{bmatrix}p_j(1) & p_j(2) & \cdots & p_j(N)\end{bmatrix}$ can be interpreted as a discrete probability distribution over features in the feature library.

    We now define useful properties and operations related to feature-concept association distributions.
    
    \subsubsection{Specificity} Specificity is the degree to which a concept has strong, specific associations with features over the feature library. For color, specificity refers to the `peakiness' of a color-concept association distribution. Concepts can have strong color associations that are concentrated in one part of color space (e.g., reds for concepts like raspberry) or divided over different parts of color space (e.g., reds and greens for watermelon) \cite{rathore2020}. Thus, we quantify specificity using \textit{entropy} of the distribution, which captures how `flat' vs. `peaky' a distribution is, regardless of how many peaks there are.
    
    \textbf{Entropy} for a feature-concept association distribution is defined as:
    \begin{equation} \label{eq:entropy}
     \vspace{-1mm}
        H_j := -\sum_{i=1}^{N} p_j(i)\log p_j(i).
         \vspace{-1mm}
    \end{equation}
    If all features in the feature library are equally associated with concept $j$, the distribution $p_j$ will be uniform, entropy will be high, and specificity will be low. If a concept $j$ is strongly associated with some features and not others, then entropy will be lower and specificity will be higher. This property of color-concept association distributions aligns with previous measures of colorability \cite{lin2013,setlur2016} (see Figure \ref{fig:entropy} in the Supplementary Material).
    
    \textbf{Mean entropy} of a concept set is the mean of the entropy of all concepts in the set: $ H_\mu := \frac{1}{n}(H_1+\dots+H_n)$.
    
    \subsubsection{Distribution difference} \label{sec:distdiff}
    We quantify distribution difference between concepts by comparing their normalized feature-concept associations.
    
    \textbf{Total variation (TV)} is what we use when comparing two concepts, say $j_1$ and $j_2$. TV is defined as follows.
    \begin{equation} \label{eq:TV}
     \vspace{-1mm}
        \text{TV}(j_1,j_2):=\frac{1}{2}\sum_{i=1}^{N} \left|p_{j_1}(i)-p_{j_2}(i)\right|.
         \vspace{-1mm}
    \end{equation}
    TV ranges between $0$ and $1$, where $\text{TV}=0$ means the two distributions are identical, and $\text{TV}=1$ means they are disjoint (for each feature $i$, either $p_{j_1}(i)$ or $p_{j_2}(i)$ must be zero).
    
    \textbf{Generalized total variation (GTV)} is a generalization of TV that we defined for cases when more than two concepts must be compared, say $j_1,\dots,j_k$. We define GTV as follows.
    \begin{equation} \label{eq:GTV}
        \text{GTV}(j_1,\dots,j_k) := -1 + \sum_{i=1}^{N} \max(p_{j_1}(i), p_{j_2}(i),\ldots,p_{j_k}(i)).
         \vspace{-1mm}
    \end{equation}
    In the case where $k=2$, GTV reduces to TV. In other words, $\text{GTV}(j_1,j_2)=\text{TV}(j_1,j_2)$. For details on the motivation behind our definition of GTV, see the Supplementary Material, Section~\ref{sec:GTV}.
    
    \subsubsection{Structure-agnostic property}
    The notions of entropy, TV, and GTV are agnostic to intrinsic structure of the feature source. For example, perceptual color spaces are structured according to perceptual similarity, but entropy of a color-concept distribution depends on the fraction of the colors that are highly associated with the concept, regardless of perceptual similarity. We chose structure-agnostic metrics for specificity and distribution difference so that semantic discriminability theory could readily generalize to feature sources with less well-defined metric spaces (e.g., shape, texture, odor).
    
    \subsection{Semantic discriminability}\label{sec:semdistthry}
    As described in Section~\ref{sec:assignment}, semantic discriminability of perceptual features is the ability to infer a unique mapping between features and concepts. It is reflected in the degree to which inferred mappings vary among individuals or within individuals between trials. We model this variability by treating feature-concept associations as random variables. Rather than solving an assignment problem using the mean $a_{ij}$ values, we look at the \textit{probability} of the likeliest assignment, where probability is computed with respect to  uncertainty in the $a_{ij}$. We now make this notion more precise.
    
\begin{figure}[ht]
     \centering
     \includegraphics[width=1.0\columnwidth]{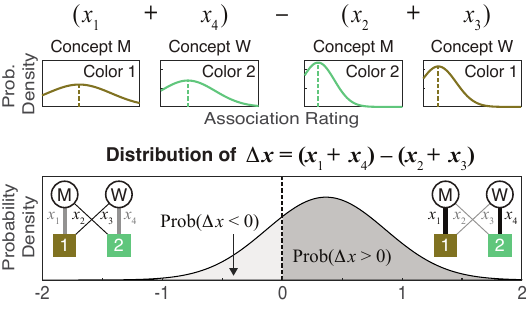}
     \caption{Diagram from~\cite{schloss2021} that shows how association ratings between concepts \{M,W\} and colors \{1,2\} produce a distribution for $\Delta x$. Semantic distance is the absolute difference of the area under the curve to the left and right of zero.}
     \label{fig:semdisteq}
     \vspace{-2mm}
    \end{figure}

    \textbf{Semantic distance} is a way to operationalize semantic discriminability in the case where there are $n=2$ features and concepts~\cite{schloss2021}. Figure~\ref{fig:semdisteq} illustrates an example in which we have concepts \{M,W\} and colors \{1,2\}. The color-concept associations between all possible pairs are $x_1,\dots, x_4$, as shown in Figure~\ref{fig:semdisteq}. We assume each $x_k$ is normally distributed with mean $\bar x_k$ equal to the corresponding $a_{ij}$ and standard deviation $\sigma_k = 1.4\cdot\bar x_k(1-\bar x_k)$, which was found to be a good fit to experimental data~\cite{schloss2021}. The outcome of the assignment problem is determined by the quantity $\Delta x := x_1-x_2-x_3+x_4$. The optimal assignment is:
    (M-1 and W-2 if $\Delta x > 0$) and (M-2 and W-1 if $\Delta x < 0$).
    Semantic distance is defined by the equation
    \begin{equation}\label{eq:sd}
        \Delta S = |\text{Prob}(\Delta x > 0)-\text{Prob}(\Delta x < 0)|.
    \end{equation}
    Since the $x_k$ are assumed to be normally distributed, so is $\Delta x$, and the probabilities in \eqref{eq:sd} can be computed analytically:
    \begin{equation}\label{eq:probpositive}
        \text{Prob}(\Delta x >0) = \Phi\left(\frac{(\overline{x}_1+\overline{x}_4)-(\overline{x}_2+\overline{x}_3)}{\sqrt{\sigma_1^2+\sigma_2^2+\sigma_3^2+\sigma_4^2}} \right),
    \end{equation}
    and $\text{Prob}(\Delta x < 0)=1-\text{Prob}(\Delta x > 0)$, where $\Phi(\cdot)$ is the cumulative distribution function (cdf) of the standard normal distribution.
    When $\Delta S$ is close to $0$, $\Delta x$ has a similar probability of being positive or negative, so the assignment is fragile. When $\Delta S$ is close to $1$, $\Delta x$ is almost always positive or almost always negative, so the assignment is robust. This notion of semantic distance can be used even when the features are not colors, by replacing the color-concept associations with feature-concept associations, and adjusting the formula for $\sigma_k$ as appropriate.
    
\begin{figure*}[ht!!]
 \centering
 \includegraphics[width=1.0\textwidth]{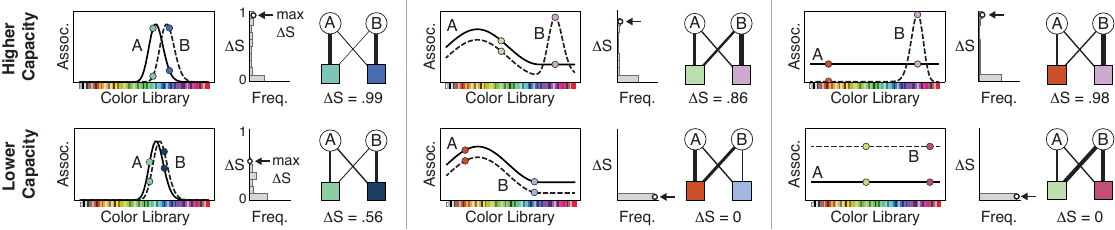}
 \caption{Hypothetical color-concept association distributions for concepts A and B, showing how capacity varies with distribution difference (top row: higher capacity; bottom row: lower capacity). In each column, the distribution for concept A is the same and concept B varies.  The histograms to the right show how distribution difference affects capacity with arrows pointing to maximum semantic distance ($\Delta S$) for the concept set. Corresponding bipartite graphs show the color set with maximum semantic distance (this is arbitrary when the distributions are parallel because semantic distances for all color pairs are equally poor).}
 \label{fig:toy_profiles_v2}
 \vspace{-4mm}
\end{figure*}

    \textbf{Generalized semantic distance} is an extension of semantic distance to the case where there are $n > 2$ features and concepts. In this case, there will be $n!$ ($n$ factorial) possible assignments. We define generalized semantic distance in a manner analogous to semantic distance; we label the feature-concept associations between all possible pairs as $x_1,x_2,\dots,x_{n^2}$ and assume they are normally distributed random variables.\footnote{\label{foot}Here, we use color-concept association ratings, so we assume the $x_k$ are distributed with the same $\sigma_k$ used to define semantic distance~\cite{schloss2021}. In principle, the distributions of the $x_k$ can be changed to suit other cases beyond color.}
    In this more complicated scenario, the assignment is not determined by a simple quantity such as $\Delta x$ and no formula analogous to \eqref{eq:probpositive} exists to determine the assignment. Instead, we use the following Monte Carlo approach.
    \begin{enumerate}
        \item Sample $x_1,\dots,x_{n^2} $ from the distribution of merit scores$^{\ref{foot}}$ and solve an assignment problem using the sampled merit scores.\label{step1}
         \vspace{-2mm}
        \item Repeat step~\ref{step1} a large number of times and count the number of times each distinct assignment occurs.  Let $p$ be the proportion of times that the most frequent assignment occurred. Since there are $n!$ possible assignments, we must have $\frac{1}{n!} \leq p \leq 1$.\label{step2}
         \vspace{-2mm}
        \item The generalized semantic distance $\Delta S$ is defined as a linear rescaling of $p$ to ensure that $0\leq \Delta S \leq 1$. The formula is:
        \begin{equation}\label{eq:gensemdist}
            \Delta S = \frac{n!p-1}{n!-1}.
            \vspace{-0mm}
        \end{equation}
        
    \end{enumerate}
    A similar Monte Carlo approach was used in~\cite{schloss2018} to predict the results of assignment inference in a recycling task (6 concepts and 6 colors).

    Just like semantic distance, generalized semantic distance is a number between $0$ and $1$, where a larger number indicates more robust assignments, and consequently, higher semantic discriminability. We use the same symbol $\Delta S$ for both notions of distance because in the case where $n=2$, generalized semantic distance is (on average) equal to semantic distance, and the approximation becomes exact as the number of samples in step~\ref{step2} tends to infinity. Conversely, in the limit $n\to\infty$, we have $\Delta S \to p$ and the rescaling in~\eqref{eq:gensemdist} has no effect.

    \textbf{Semantic contrast} is similar to generalized semantic distance, except it estimates the proportion of times a given color is assigned to the ``optimal'' concept (compared to all other assignments). This estimation is computed using the Monte Carlo method described earlier, with  optimal defined by the solution to an assignment problem using the balanced merit function computed on feature-concept associations. 
 

For a given concept, the optimal color for that concept may have higher semantic contrast in one context and lower semantic contrast in another context, depending on the other colors and concepts in the encoding system. A concept set that has higher capacity for semantic discriminability (Section \ref{sec:cap}) should enable higher semantic contrasts among colors in its optimal palette.


    
    The steps to computing semantic contrast are:
    (1) Solve an assignment problem (see Section \ref{sec:assignment}) using the mean association ratings $\bar x_1,\dots,\bar x_{n^2}$. We call this the \textit{optimal assignment}.
    (2) Sample $x_1,\dots,x_{n^2}$ from the distribution of merit scores and solve an assignment problem using the sampled merit scores.
    (3) Repeat step 2 a large number of times and count the proportion of times each feature was assigned to the same concept as in the optimal assignment 
    This proportion is each feature's semantic contrast.

\subsection{Capacity for semantic discriminability} \label{sec:cap}
Capacity for semantic discriminability is the extent to which it is possible to produce semantically discriminable features for a given set of concepts. We operationalized capacity for semantic discriminability (\textit{capacity} for short), using \textbf{max capacity}. This is a scalable measure that returns the semantic distance of the most semantically discriminable feature set for a concept set, given a feature library. 

To compute max capacity for a given concept set, we solve an assignment problem using the balanced merit function (Section~\ref{sec:assignment}) over the entire feature library. This yields a feature set. We define max capacity as the (generalized) semantic distance of this feature set for the given concept set. High max capacity indicates that the feature library contains at least one feature set with high semantic discriminability for the concept set. Low max capacity indicates no such feature set exists for that concept set, at least given the feature library. 

In the case of two concepts, the balanced merit approach for computing max capacity gives the same result as exhaustively computing the semantic distance for each pair of colors, then finding the maximum of those semantic distances. Using balanced merit, though, allows max capacity to scale easily; it can be efficiently computed for large concept sets and feature sets.
We also explored alternative ways to operationalize capacity (see Supplementary Material Section~\ref{sec:semdisc_capacity}).

\subsection{The theory}

\textbf{Semantic discriminability theory} posits that the capacity to produce semantically discriminable perceptual features for a set of concepts depends on the difference in feature-concept association distributions over a feature library. Briefly, distribution difference predicts capacity, distinct from the contribution of specificity. This idea differs from previous approaches, which primarily focused on color-concept associations for concepts in isolation when evaluating the potential to meaningfully encode particular concepts using color \cite{lin2013, setlur2016}. 

Figure \ref{fig:assoc} shows the distinction between distribution difference and specificity of color-concept associations, with respect to capacity. It includes concept sets with large, medium, and small distribution differences. Capacity is illustrated with histograms below each concept set. They show the frequency of color sets across values of semantic distance (2485 possible 2-color sets from the UW-71 color library), with an arrow pointing at maximum semantic distance. Concept sets with large, medium, and small distribution differences result in high, medium, and low capacity, respectively. Yet, the concepts with medium capacity (driving and comfort) have far lower specificity than concepts with low capacity (eggplant and grape). The reason that concepts with low specificity can result in higher capacity than concepts with high specificity is that semantic discriminability depends on the difference in merit of each possible set of feature-concept assignments, not just isolated feature-concept associations (Section \ref{sec:assignment}). 

 Figure~\ref{fig:toy_profiles_v2} further illustrates this point with hypothetical color-concept association distributions for 2-concept sets that have higher capacity (top row) and lower capacity (bottom row). The colored dots on the distributions indicate the optimal assignment according to balanced merit (though this is arbitrary when the distributions are parallel because all assignments are equally poor). Next to each distribution pair is a histogram of semantic distances (as in Figure \ref{fig:assoc}) and a bipartite graph for the colors with maximum semantic distance (thicker edges connecting colors and concepts indicate greater merit). Semantic distance ($\Delta S$) is indicated below the bipartite graphs, and can be visually inspected by comparing the total merit of the outer edges vs. inner edges and assessing the degree to which one sum is larger. When distribution difference is high (top row), capacity is high, even if one concept has a uniform distribution (i.e., no specificity). However, when distribution difference is lower (bottom row), capacity is lower, even if both concepts have high specificity. 

We chose the particular examples in Figure \ref{fig:assoc} and Figure \ref{fig:toy_profiles_v2} to highlight the dissociation between distribution difference and specificity, but we systematically tested for effects of these factors on capacity in Experiment 1.

\section{Experiment 1}\label{sec:exp1}
Experiment 1 tested the hypothesis that capacity for semantic discriminability is predicted by distribution difference, independent of specificity. We first collected color-concept association data from human participants, and used those data to calculate capacity, distribution difference, and specificity. We then tested our hypothesis on 2-concept sets (Section \ref{sec:set2}) and 4-concepts sets (Section \ref{sec:set4}). Semantic discriminability predicts people's ability to interpret palettes in visualizations \cite{schloss2021}, so our modeling approach for understanding capacity for semantic discriminability should have implications for interpretability. 
The code and data for all experiments is at: \url{https://github.com/SchlossVRL/sem_disc_theory}.

\subsection{Methods}

\subsubsection{Participants}
185 undergraduates participated for credit in Introductory Psychology (mean age =18.66, 99~females, 86~males, gender provided through free-response). All gave informed consent and the UW--Madison IRB approved the protocol. Color vision was assessed by asking participants if they had difficulty distinguishing between colors relative to the average person and if they considered themselves colorblind. Participants were excluded if they answered yes to either (5 excluded).

\subsubsection{Design, Displays, and Procedure} 
Participants judged the association between each of 71 colors and each of 20 concepts. The colors were the UW-71 color library, an extension of the UW-58 colors \cite{schloss2021}, see Supplementary Material for details and Table \ref{table:UW_71_colors} for CIELAB coordinates.\footnote {We converted CIELAB to RGB using MATLAB's \lstinline{lab2rgb} function, which makes assumptions about monitor characteristics, so the colors were not exact renderings of CIELAB coordinates. Without calibration, the colors rendered by RGB coordintes may vary across monitors, but using a fixed correspondence between D65 CIELAB and RGB can approximate intended colors online \cite{stone2014}.} The concepts were from Lin et al. \cite{lin2013}, including 5 concepts in each of four concept categories (fruits, vegetables, activities, and properties) (Table \ref{tab:concepts}). Participants were randomly assigned to one of four categories (fruits $n=46$, vegetables $n=45$, activities $n=45$, properties $n=44$). They judged all colors for all five concepts within their assigned category (71  colors $\times$ 5 concepts $=$ 355 trials). Trials were presented in a blocked randomized design---all colors were presented in a random order for a given concept before starting the next concept, and concept order was also randomized. 

The displays included the concept word centered at the top of the screen (font-size: 24 pt, font-family: Lato) and colored square centered below (80~px $\times$ 80~px). Below the colored square, was a line-mark slider scale (400 px long), with the left end labeled ``not at all'' and the right end labeled ``very much'' and the center marked with a vertical line (3 px wide and 32 px tall). The background was gray (CIE Illuminant D65, x = .3127, y = .3290, Y = $10$~cd/$\text{m}^2$), so that very dark colors (e.g., black) and very light colors (e.g., white) could be seen against the background. Data were recorded in pixel units, and scaled to range from 0-1. Displays were generated using the jsPsych JavaScript library \cite{de2015jspsych}, presented on participants' personal devices.

Participants were told they would see a set of concepts and series of colors, one concept and color at a time. Their task was to rate how much they associated the color with the concept by moving the slider on the scale from ``not at all'' to ``very much'', and clicking ``next'' to continue. Before beginning, they were shown a list of all concepts and  the UW-71 colors. They were asked to anchor the endpoints of the rating scale for each concept \cite{palmer2013} by thinking about which color they associated the most/least with that concept, and considering these colors as representing the ends of the slider scale for that concept. During the experiment, ratings were blocked by concept, and after each block participants were told how many blocks remained.

 \begin{table}
  \caption{Full set of concepts in Experiment 1 (first four columns of concepts were used in Experiment 2).}
  \label{tab:concepts}
 
  \begin{tabular}{p{1.4 cm}|p{1cm}p{.7cm}p{.9cm}p{.9cm}l}  
  \toprule
  \bf{Category} &&&  \bf{Concepts} \\
  \midrule
  \bf{Fruits} & peach & cherry & grape & banana & apple \\
  \bf{Vegetables} & corn & carrot & eggplant & celery & mushroom \\
  \bf{Activities} & working & leisure & sleeping & driving & eating \\
  \bf{Properties} &efficiency & speed & safety & comfort & reliability \\
  \bottomrule
  \end{tabular}
  \vspace{-0mm}
\end{table}

\subsection{Results and Discussion}

\subsubsection{2-Concept sets} \label{sec:set2}
We began by calculating the mean color-concept association ratings over participants. Next, for all $k=2$ concepts out of the $n = 20$ concepts in Table \ref{tab:concepts} (190 2-concept sets in total), we used the mean color-concept associations to calculate capacity for semantic discriminability, distribution difference, and mean specificity. To calculate capacity, we followed the method in Section \ref{sec:cap}. To calculate distribution difference, we used total variation (TV) in Equation \eqref{eq:TV} and normalized the TV values to range from 0 to 1. To calculate mean specificity, we first computed entropy ($H$) for each concept (Equation \eqref{eq:entropy}) over $N=71$ colors, and then computed the mean entropy over concepts within each set. Given that higher specificity corresponds to lower entropy, we normalized mean entropy to range from 0 to 1 and subtracted the scores from 1, such that larger numbers indicated higher specificity. Figure \ref{fig:entropy} in the Supplementary Material shows the raw entropy for each concept. Concepts with lower entropy/higher specificity corresponded to colorable concepts in \cite{lin2013}, and concepts with higher entropy/lower specificity corresponded to non-colorable concepts in \cite{lin2013}. 

\begin{figure}[t]
 \centering
 \includegraphics[width=1.0\columnwidth]{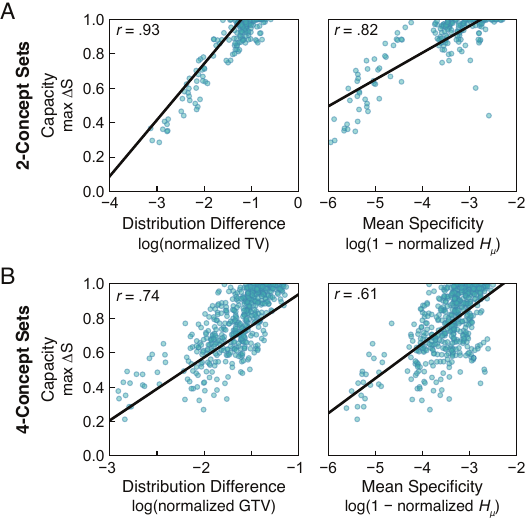}
 \caption{Relations between capacity for semantic discriminability and distribution difference (log(normalized (generalized) total variation distance); left) and specificity (log($1-$ normalized mean entropy); right) for 2-concept sets (top) and 4-concept sets (bottom). For 4-concept sets we downsampled from 4845 points to 500 points to avoid overplotting.}
 \label{fig:scatter}
    \vspace{-0mm}
\end{figure}

Figure \ref{fig:scatter}A shows the relation between capacity for semantic discriminability and distribution difference (left), and mean specificity (right). For both distribution difference and mean specificity, we plotted the log of the normalized scores to preserve linearity.  The correlation between capacity and distribution difference over all 190 2-concept sets was strongly positive ($r(188)=.93$, $p<.001$), with a strong trend for capacity to increase with increased distribution difference. The correlation between capacity and mean specificity was also significantly positive ($r(188)=.82$, $p<.001$), but was significantly weaker than the correlation with distribution difference (Fisher's r-to-z transformation $z(188)=4.85$, $p<.001$). This weaker correlation can be attributed, in part, to there being concept sets with high capacity, despite moderate to low  mean specificity, and concept sets with low capacity despite high mean specificity (Figure \ref{fig:scatter}, right).

To examine whether distribution difference and mean specificity contributed independently to capacity, we used a multiple linear regression model to predict capacity from these two factors (z-scored to center them and put them on the same scale). As shown in Table \ref{tab:Exp1and2reg}, distribution difference was a strong significant predictor, and mean specificity was not significant. Thus, the variance explained in capacity by distribution difference was independent from mean specificity, and mean specificity did not contribute after accounting for distribution difference.

\subsubsection{4-Concept sets} \label{sec:set4}


For all $k=4$ concepts out of the $n = 20$ concepts in Table \ref{tab:concepts} (4845 4-concept sets in total), we used the mean color-concept associations to calculate capacity, distribution difference, and mean specificity, as described in Section \ref{sec:set2} for 2-concept sets. However, instead of semantic distance to compute capacity we used generalized semantic distance (Section \ref{sec:semdistthry}), and instead of using TV to compute distribution difference, we used GTV (Equation \ref{eq:GTV}, Section \ref{sec:distdiff}). 

Figure \ref{fig:scatter}B shows the relation between capacity for semantic discriminability and distribution difference (left), and mean specificity (right) for 4-concept sets. As for 2-concept sets, we used the log of the normalized distribution difference and mean specificity scores to preserve linearity. Capacity was positively correlated with both distribution difference ($r(4843)=.74$, $p<.001$) and mean specificity ($r(4843)=.61$, $p<.001$), but the correlation with distribution difference was  greater (Fisher's r-to-z transformation ($z(4843)=11.88$, $p<.001$).


Using the same regression analysis as for 2-concept sets, distribution difference was a strong significant predictor (Table \ref{tab:Exp1and2reg}). Mean specificity a weak significant predictor, but surprisingly it was negative, such that less specificity resulted in greater capacity in the context of this model. 

  \begin{table}
  \caption{Multiple linear regression predicting capacity for semantic discriminability from distribution difference and mean specificity for all 2-concept sets and 4-concept sets.}
  \label{tab:Exp1and2reg}
  \setlength\tabcolsep{5 pt} 
  \begin{tabular}{clrrrr}  
  \toprule
  \bf{Model} & \bf{Factor} & $\boldsymbol{\beta}$ & \bf{SE} & $\boldsymbol{t}$ & $\boldsymbol{p}$ 
  \\
  \midrule
  2-concept & Intercept & $.867$ & $.005$ & $181.9$ & $<.001$
  \\ 
  & Distribution diff. & $.160$ & $.010$ & $15.6$ & $<.001$
  \\ 
  & Specificity & $.002$ & $.010$ & $.201$ & $.841$
  \\ 
  \midrule
  4-concept & Intercept & $.772$ & $.002$ & $483.8$ & $<.001$
  \\
  & Distribution diff. & $.235$ & $.004$ & $53.6$ & $<.001$
  \\
  & Specificity & $-.112$ & $.004$ & $-25.5$ & $<.001$
  \\
  \bottomrule
  \end{tabular}
  \vspace{-4mm}
\end{table}

In summary, Experiment 1 supports the hypothesis that the capacity to produce semantically discriminable color palettes for a set of concepts depends on the difference in color-concept association distributions, independent of specificity. Considering specificity of color-concept associations in isolation is insufficient. These results emphasize the importance of considering relative color-concept associations for a given set of concepts, rather than the concepts in isolation, when evaluating the potential for semantically discriminable color palettes.


\section{Experiment 2}\label{sec:exp2}

Semantic discriminability theory implies that if concept sets have high capacity for semantic discriminability, it should be possible to create encoding systems assigning those concepts to colors that people can interpret. People should be able to interpret the correct mappings between colors and concepts, even for concepts previously considered ``non-colorable,'' insofar as the colors are semantically discriminable. We tested this hypothesis in Experiment 2. We defined accuracy as the proportion of responses that matched the optimal mapping specified by an assignment problem using the balanced merit function (see Section \ref{sec:SC_supp} in the Supplementary Material for a further discussion on accuracy, and its relation to measures of semantic discriminability).



 \subsection{Methods}
 
\subsubsection{Participants} 
98 participants (74 males, 24 females) were recruited on Amazon Mechanical Turk. All gave informed consent, and the UW--Madison IRB approved the protocol. Eight were excluded for not reaching 100\% accuracy on catch trials (Section \ref{sec:e3design}), three of which reported atypical color vision. All other participants reported typical color vision.
    
\subsubsection{Design, Displays, and Procedure} \label{sec:e3design} 

\begin{figure}[ht]
 \centering
 \includegraphics[width=1.0\columnwidth]{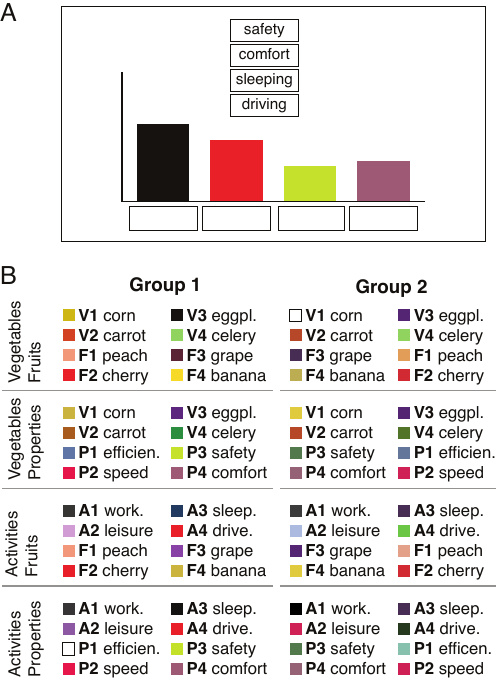}
 \caption{(A) Example trial in Experiment 2. Participants labeled each bar by clicking the label and dragging/dropping it in the box below the bar. (B) Palettes and corresponding concepts used to construct the stimuli (see text for details).}
 \label{fig:allpals}
 \vspace{-3 mm}
\end{figure}

For each trial, participants were presented with a bar graph centered on the screen, consisting of four colored bars (Figure \ref{fig:allpals}A). Each bar was 130~px wide and varied in height randomly (from 260-300 px high). The bars were spaced 45~px apart. At the start of the trial, a set of four concepts (22 pt font) was centered above the graph in a random order. The y-axis was unlabeled. Below the x-axis, there were empty boxes 120-px wide and 50~px high. During the trial, participants labeled each bar by clicking on a label and dragging/dropping it in the empty box below the bar. 
The displays were generated using the Charts.js and jsPsych JavaScript libraries.

Each participant completed 64 trials, which included 8 color-concept sets $\times$ 8 color positionings within each set. Figure \ref{fig:allpals}B shows the palettes for each set. The stimuli were constructed using displays like in Figure \ref{fig:allpals}A, but swapping out the concept sets and corresponding color palettes, and balancing the bar color positioning as follows. 

\textbf{Concept sets.} To generate the concept sets, we randomly selected four concepts from each of the concept categories from Experiment 1 (fruits (F), vegetables (V), activities (A), properties (P)) and labeled them 1-4 (Table \ref{tab:concepts}, Figure \ref{fig:allpals}B). We then tied pairs of concepts within each category (e.g., V1-V2, V3-V4). We combined pairs of concepts such that all participants saw (1) vegetables with fruits, (2) vegetables with properties, (3) activities with fruits, and (4) activities with properties. Using this design, we created two groups of stimuli, divided over two groups of participants to reduce the number of trials for any one participant. Group 1 saw sets of four concepts, with concepts 1 and 2 in one category paired with concepts 1 and 2 in the other category (e.g., V1-V2 with F1-F2), and sets of four concepts with concept 3 and 4 in one category paired with concepts 3 and 4 in the other category (e.g., V3-V4 with F3-F4). Group 2 saw the opposite pairings (e.g., V1-V2 with F3-F4, V3-V4 with F1-F2). Within this design, all participants saw each concept an equal number of times. Participants were randomly assigned to Group 1 ($n = 47$) or Group 2 ($n = 43$).

\textbf{Color palettes.} For each concept set, we generated its color palette using the balanced merit function (Equation \eqref{eq:balanced_merit}) in an assignment problem. The resulting assignments determined the encoded mapping we defined as ``correct.'' We used the balance merit function because previous evidence suggested it aligns with the merit people use in assignment inference (see Section \ref{sec:assignment}). Merit was computed over the color-concept association data reported in Experiment 1 for all 71 colors in the UW-71 library \footnote{Due to a scaling issue during palette creation, 15 of the 64 color-concept pairings were not optimal. This did not affect the analyses, but accuracy may have been greater if participants had seen fully optimized palettes.} The color palettes are shown in Figure \ref{fig:allpals}B. The CIELAB coordinates for the palette colors can be found at \url{https://github.com/SchlossVRL/sem_disc_theory}. The graphs were presented on a gray background approximating CIE Illuminant D65 ($x = .3127$, $y = .3290$, $Y =10$, cd/m$^2$).

\textbf{Bar color positioning.} 
Each of the eight color-concept sets for a given group (Figure \ref{fig:allpals}B) was presented eight times in eight bar color positionings along the x-axis. This was done using a blocked randomized design, so all eight color-concept sets appeared once in a random order, randomly assigned to a color positioning within a block, before starting the next block. The eight possible color positionings were defined using a Latin square design (four positionings, left/right reversed). Thus, within a color set, each color appeared in each of four positions twice, with the colors to its left/right in opposite positions.

\textbf{Catch trials.} We included eight catch trials, one per block, in which bars were colored a shade of red, yellow, green, and blue, and the labels were ``red'', ``yellow'', ``green'', and ``blue.'' We set an \textit{a priori} exclusion criterion that participants must be 100\% accurate on these catch trials, otherwise their data would be excluded from analysis.   

Participants were told they would see a series of colored bar graphs, with four bars and four words at the top of the screen. Their task was to match each word to its corresponding bar color by clicking and dragging the label to the empty box below the bar. They were told to use their best guess if they were unsure how to match the labels to the bar colors. They then completed a practice trial with four concepts that were not in the main experiment (blueberry, mango, strawberry, lemon) and colors chosen by the balanced merit function. Associations for these concepts had been collected for a different project. During the trials, all bars had to be labeled before a ``continue'' button could be pressed to go to the next trial. Once placed in a box, a label could be dragged to another box and all labels could be reset to the starting position by pressing a  ``reset label'' button. Trials were separated by a 100 ms. inter-trial interval. Participants received breaks after each block, and were told the proportion of completed trials at each break.

 \begin{figure*}[h!!]
 \centering
 \includegraphics[width=1.0\textwidth]{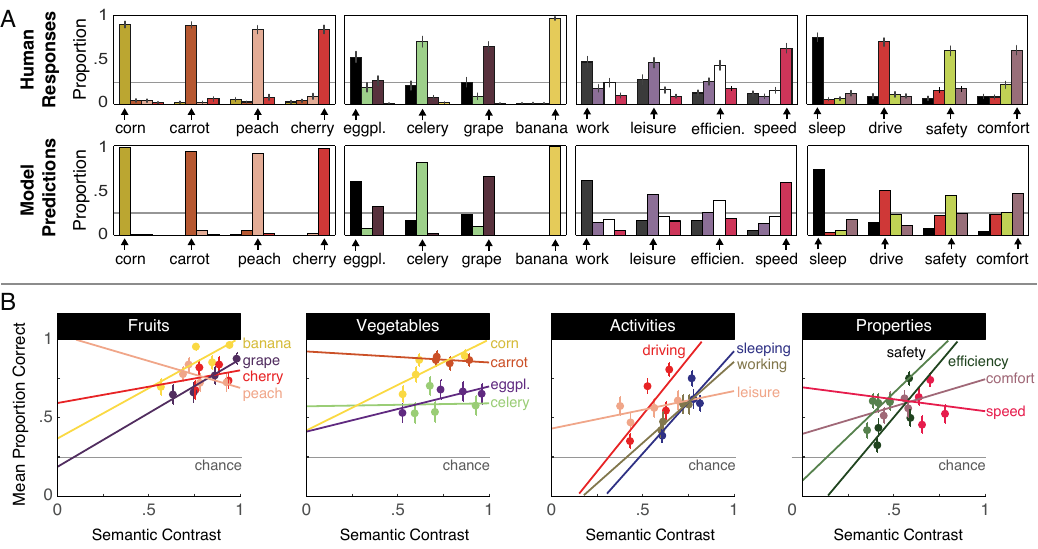}
 \caption{(A) Proportion of times participants chose each color (top) and predicted proportions from generalized semantic distance (bottom) for a subset of palettes from group 1 (see Figure \ref{fig:modelVdataALL} in the Supplementary Material for the full dataset). The correct response for each concept is marked by an arrow along the x-axis. The colors of the bars correspond to the colors of the stimuli. (B) Mean proportion correct for each concept in each palette as a function of semantic contrast of its correct color (best fit lines drawn for each concept). All the points for a given concept and corresponding best fit line are shown in the same color to help group the points in this figure (these colors were not necessarily the colors shown in the experiment. In (A) and (B) gray horizontal lines correspond to chance (.25) and error bars represent ± standard errors of the means. }
 \label{fig:exp2results}
 \vspace{-4mm}
\end{figure*}

 \subsection{Results and Discussion}
For each participant, we calculated the proportion of times they chose each concept for each color in each color-concept set, averaged over bar color positioning. These results are shown for a subset of the concept sets in Figure \ref{fig:exp2results}A (top row), and for all concepts sets in Figure \ref{fig:modelVdataALL}. 
For each color-concept pairing, we calculated accuracy as the proportion of trials in which participants selected the optimal pairing (defined with respect to balanced merit) (Section \ref{sec:SC_supp}). The arrows below the x-axis in Figures \ref{fig:exp2results}A and \ref{fig:modelVdataALL} point up to the correct color.

 \begin{table} 
  \caption{Logistic mixed-effect model predicting accuracy from specificity of the concept, semantic contrast of the concept's correct color, and association between the concept and its correct color.}
  \label{tab:exp3}
  \begin{tabular}{lrrrr}  
  \toprule
 \bf{Fixed Effects} & $\pmb{\beta}$ & \bf{SE} & $\pmb{z}$ & $\pmb{p}$ 
  \\
  \midrule
Intercept & $1.272$ & $.176$ & $7.249$ & $<.001$
  \\ 
   Specificity & $.226$ & $.088$ & $2.577$ & $.001$
  \\ 
   Semantic Contrast & $.645$ & $.081$ & $7.926$ & $<.001$
  \\ 
  Association Strength & $-.064$ & $.057$ & $-1.12$ &$.262$
  \\
  \bottomrule
  \end{tabular}
  \vspace{-4mm}
\end{table}

We first tested whether concept sets with higher capacity enabled creating encoding systems that were easier to interpret. To do so, we correlated max capacity for each of the 16 concept sets with mean accuracy over all colors within each set. There was a significant relation ($r = .58, p < .02$), indicating greater capacity for semantic discriminability corresponded to greater interpretability. 

Next, we tested whether participants' patterns of color choices for each concept were correlated with model predictions computed by solving an assignment problem with perturbed association ratings (the Monte Carlo process described in Section \ref{sec:semdistthry} over 1000 iterations). These predictions are shown in the bottom row of Figure \ref{fig:exp2results}A and in Figure \ref{fig:modelVdataALL}. In the model predictions, the height of the bars correspond to the proportion of times each color was assigned to each concept. The predictions strongly correlated with participant responses over the full dataset of 4 colors $\times$ 16 4-concept sets ($r(126) = .95, p < .001$), with high correlations for each group (Group~1: $r(126) = .96$, Group~2: $r(126) = .94, ps < .001$). 



Finally, we tested our hypothesis that participants would be able to interpret the correct mappings between individual colors and concepts, insofar as the colors were semantically discriminable. Figure \ref{fig:exp2results}A shows that participants chose the correct colors well above chance, even for concept sets in which all concepts have been called non-colorable (e.g., $\{$sleeping, driving, safety, speed$\}$. To examine whether accuracy for given a concept varied depending on semantic discriminability of its correct color, in \ref{fig:exp2results}B, we plotted accuracy for each concept as a function of the semantic contrast of its correct color (see Section \ref{sec:semdistthry} and Section \ref{sec:SC_supp} for details on semantic contrast). Plots are separated by concept category, with four points per concept, corresponding to the four color-concept sets in which it appeared. Generally, the slopes of the best fit lines for each concept were positive, indicating that accuracy increased with semantic contrast. Responses for some concepts (e.g., fruits) were highly accurate for all color-concept sets because their optimal colors have high semantic contrast in all concept sets we tested. 

We analyzed this pattern of accuracies using a mixed-effect logistic regression model predicting accuracy for each concept in each set using three factors: semantic contrast of the correct color for that concept (relative to the other colors in the palette), specificity of the concept as defined in Experiment 1, and association strength between the concept and its correct color (previously shown to influence accuracy in similar tasks \cite{schloss2018, schloss2021}). These predictors were calculated using data from Experiment 1 (different participants from Experiment 2). We also included by-subject random intercepts and by-subject random slopes for each factor. We z-scored the individual predictors to put them on the same scale and set the correlations between the random slopes to be 0 to help the model converge. As shown in Table \ref{tab:exp3}, accuracy significantly increased with greater semantic contrast and with greater specificity. Association strength was not significant. 


 Overall, accuracy was greater for concepts previously considered colorable (fruits and vegetables) ($M = 0.76$, $SD = 0.23$) than those considered non-colorable (activities and properties) ($M =0.56$, $SD = 0.24$) (Figure \ref{fig:exp2results}B). But, all activities and properties had at least one instance that was as accurate as fruits and vegetables, and all instances were above chance. Moreover, accuracy for a given concept varied based on semantic contrast with its correct color, which cannot be explained by specificity of the concept in isolation. These results suggest that any concept has potential to be meaningfully encoded using color if the color has sufficient semantic contrast with other colors in the palette.

\vspace{-2mm}
\section{General Discussion and Conclusion}\label{sec:discussion}

In this paper we presented semantic discriminability theory to specify constraints on producing semantically discriminable perceptual features for visual communication. The theory states that capacity for creating semantically discriminable features for a concept set is constrained by the difference in feature-concept association distributions for those concepts. Supporting the theory, Experiment 1 showed that distribution difference between color-concept association distributions predicted capacity for semantic discriminability in 2- and 4-concept sets, independent of specificity. And, Experiment 2 indicated people can correctly interpret mappings for concepts previously considered non-colorable, but their ability to do so depended on semantic contrast with respect to the other colors in the encoding system.

Semantic discriminability theory is rooted in feature-concept associations, which can vary cross-culturally \cite{jonauskaite2019sun, jonauskaite2019machine, tham2020systematic}. The theory implies that distribution difference will predict capacity for semantic discriminability in different cultures, as long as the association distribution data reflect the associations held by a given culture.

%



The theory further implies that any factor that influences distribution difference for a set of concepts can affect capacity. Below, we propose criteria for producing distribution differences that support adequate capacity for semantic discriminability. Evaluating these criteria will help guide future work on the potential and limitations of semantic discriminability for colors and for other perceptual features. 

\textbf{Criterion 1: Need for \textit{some} specificity.} At least \textit{some} concepts in the concept set must have association distributions with some specificity. If all concepts in a set have uniform distributions, there will be no capacity for semantic discriminability (Figure \ref{fig:toy_profiles_v2}). Some perceptual features may not support specificity as well as color does, such as line orientation. If so, such features might be less useful for communicating meaning in information visualizations.

\textbf{Criterion 2: Need for feature library variability}. To be sensitive to differences in feature-concept associations, if they exist, the feature library must be sufficiently variable. In color, variability is achieved by sampling widely over color space, as opposed to sampling say, only the bluish part of the space. One can systematically sample over color spaces because color spaces are well-defined feature sources. But, such sampling may pose a challenge for less well-specified feature sources (e.g., all possible shapes or all possible textures). 


\textbf{Criterion 3: Need for large enough feature library.} The feature library must be large enough to detect small, but important differences between feature-concept association distributions. E.g., a library with only two colors, a blue and red, might be large enough to produce distinct association distributions for the concepts sky and rose, but a library with more colors (e.g., more shades of blue) would be needed to produce distinct distributions for concepts like noon sky and night sky. 

\textbf{Conclusion.} We presented and evaluated semantic discriminability theory to define constraints on creating semantically discriminable features for perceptual encoding systems. The theory implies that any concept has potential to be meaningfully encoded using color, if the criteria above are met. Thus a concept that has low specificity (i.e., uniform distribution), can meaningfully be encoded by a color, if other concepts in the set have sufficiently different distributions. This is possible because people infer globally optimal mappings between colors and concepts, even if that means inferring concepts map to weakly-associated colors. The theory implies, and our results suggest, color is more robust for visual communication than previously thought.

\acknowledgments{
We thank Rob Nowak, Melissa Schoenlein, Kevin Lande, Tim Rogers, Chris Thorstenson, Anna Bartel, and Maureen Stone for helpful discussions. This project was supported by the UW--Madison Office of the Vice Chancellor for Research and Graduate Education, Wisconsin Alumni Research Foundation, and NSF (BCS-1945303 to KBS).}


\clearpage

\bibliographystyle{abbrv}

\bibliography{Colorability.bib}

\clearpage
\appendix
\renewcommand*{\thesection}{S}
\counterwithin{figure}{section}
\counterwithin{table}{section}
\section{Supplementary Material}\label{sec:supplementary}

\subsection{UW-71 Colors} 
\begin{figure}[ht!]
 \centering
 \includegraphics[width=1.0\columnwidth]{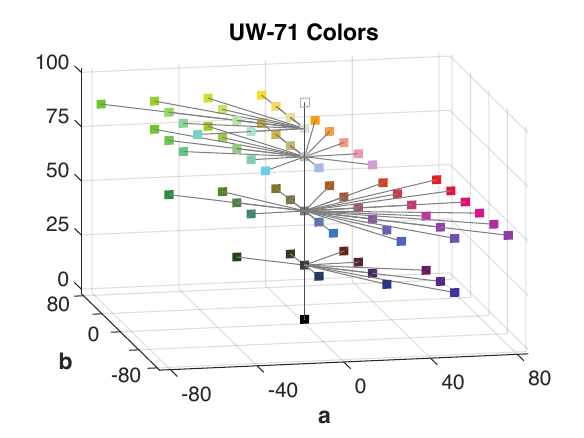}
 \vspace{-4mm}
 \caption{UW-71 colors plotted in CIELAB space.}
 \label{fig:UW71}
\end{figure}

In this study, we used a color library called the UW-71 (Figure \ref{fig:UW71}). It is based on the UW-58 colors used in \cite{rathore2020, schloss2021}, but extends to include lighter yellows and greens. The UW-58 colors includes 58 colors uniformly sampled over CIELAB space (edge distance of $\Delta$E = 25, rotated of axis by 3 degrees to increase the number of colors, as described in \cite{rathore2020}). To obtain the UW-71, we sampled an additional plane of colors at lightness $L=88$ with $\Delta E$ = 25. This provided 13 more colors, shown in the top plane of Figure \ref{fig:UW71}.
 
\subsection{Color-concept associations}

\begin{figure*}[!ht]
 \centering
 \includegraphics[width=1.0\textwidth]{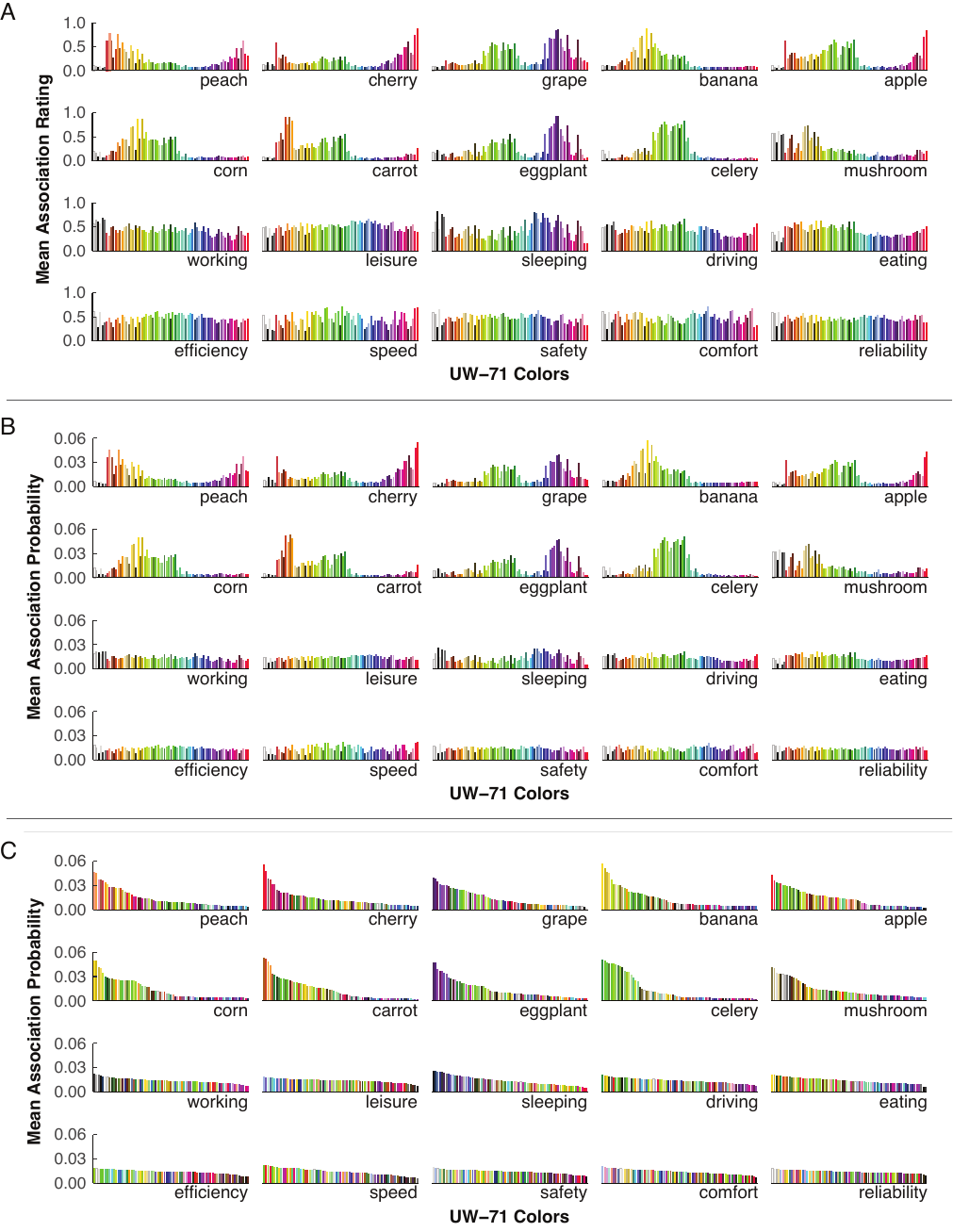}
 \vspace{-2mm}
 \caption{(A) Mean color-concept association ratings for all 20 concepts and the UW-71 colors. Each row corresponds to a different concept category (fruits, vegetables, activities, features). (B) Color-concept probability distributions as described in \ref{sec:theory}. The heights of the bars sum to 1 for each concept. (C) Color-concept probability distributions with colors sorted from high to low probability.}
 \label{fig:all_concepts}
\end{figure*}

Figure \ref{fig:all_concepts}A shows the mean color-concept association ratings collected in Experiment 1 (Section \ref{sec:exp1}). Participants rated how much they associated each color from the UW-71 color library with each of 20 concepts. In this figure, the order of the UW-71 colors along the x-axis was obtained by sorting the colors according to CIE LCh hue angle with achromatic colors being placed at the beginning of the list. Figures \ref{fig:all_concepts}B shows a transformation of the data from Figure \ref{fig:all_concepts}A, turning mean ratings for each concept into a discrete probability distribution. This was done following the normalization process described in Section \ref{sec:dists}. When we calculated entropy and (Generalized) Total Variation, we used data in this distribution form. Figure \ref{fig:all_concepts}C shows the same data from Figure \ref{fig:all_concepts}B, sorted from high to low probability within each concept. In this sorted format, it is easier see how color-concept associations vary in specificity for different concepts. We quantify specificity using the entropy of the distribution, as described in Section \ref{sec:dists}.

\subsection{Operationalizing specificity using entropy}

\begin{figure}[!ht]
 \centering
 \includegraphics[width=1.0\columnwidth]{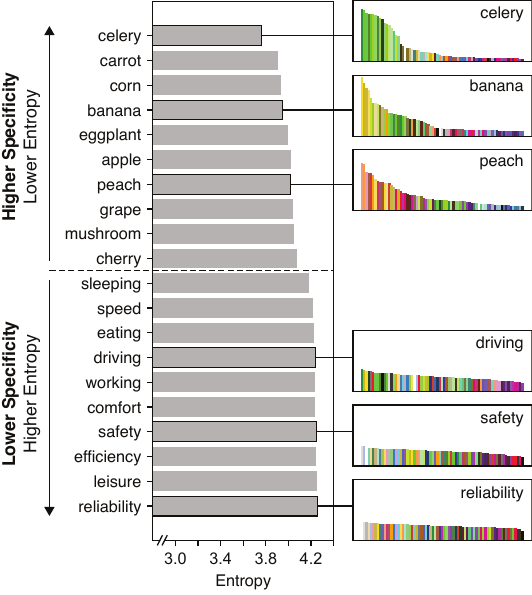}
 \vspace{-4mm}
 \caption{Entropy over color-concept association distributions (left) computed over the UW-71 color library, and examples of low- and high-entropy normed color-concept association distributions with color sorted from high to low association (right). All of the concepts above the dashed line were considered colorable and the concepts below the dashed line were considered non-colorable in \cite{lin2013}.}
 \label{fig:entropy}
\end{figure}

Figure \ref{fig:entropy} shows the entropy for each concept, computed on the normalized color-concept association distributions (Figure \ref{fig:all_concepts} B). The concepts are ordered from  high specificity (low entropy) to low specificity (high entropy), with example sorted association distributions at varying levels of specificity. The top half of the concepts with high specificity correspond to concepts that Lin et al. \cite{lin2013} called ``colorable,'' and the bottom half with low specificity correspond to concepts that they called ``non-colorable''. 

Concepts in  Figure \ref{fig:all_concepts}C that have a steeper slope of descending mean association probabilities have lower entropy and higher specificity. If one looks at the corresponding plots for each concept in Figure~\ref{fig:all_concepts}B and entropy scores in Figure~\ref{fig:entropy}, it can be seen that a concept can have multiple peaks at different locations (such as eggplant) or a single peak (such as celery) and still have similar entropy (similar slopes of ordered mean association probabilities). This comparison highlights how entropy is agnostic to any arbitrary ordering of colors and captures the 'peakiness' or 'flatness' of the distributions.

\subsection{Operationalizing capacity}\label{sec:semdisc_capacity}
We considered four ways of operationalizing capacity for semantic discriminability: (1) Maximum semantic distance (computationally identical to maximum capacity), (2) Mean semantic distance, (3) Median semantic distance, and (4) Proportion of semantic distances greater than some threshold (\emph{threshold capacity}). For 2-concept sets, we computed all possible semantic distances for each possible color set (from the UW-71 library) within each concept set. We then compared $\text{mean}(\Delta S)$, $\text{median}(\Delta S)$, and max capacity by computing the correlations between each pair of metrics for all possible feature sets within all possible concept sets. Each correlation was strong (max and mean: $r(188) = .86$,$p<.001$; max and median: $r(188) = .78$, $p<.001$; mean and median: $r(188) = .98$, $p<.001$), so those three metrics were similar.



\begin{figure}[!ht]
 \centering
 \includegraphics[width=1.0\columnwidth]{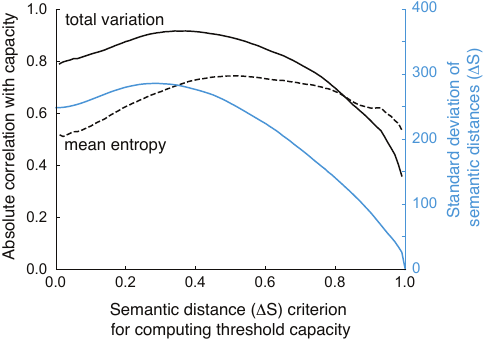}
 \caption{The effect of varying the threshold for semantic distance on non-signed correlations between total variation and threshold capacity (solid black) and mean entropy and capacity for semantic discriminability (dashed black).
 The blue line shows the standard deviation of semantic distances greater than the threshold changes as a function of the threshold. When the semantic distance criterion is 0 or 1, both of the correlations are undefined (there is no variability in capacity).}
 \label{fig:droopy}
\end{figure}

To assess threshold capacity, it is necessary to define a value of $\Delta S$ to be used as the threshold. Figure \ref{fig:droopy} shows the relation between the choice of threshold value for threshold capacity and the correlations between the resulting threshold capacity and TV (solid line) and mean entropy of the concept set (dashed line). The blue line shows that the standard deviations of the $\Delta$S values decrease as threshold increases. This decrease is expected because the average number of $\Delta$S values greater than some threshold will decrease the higher the threshold is. When the criterion for threshold is very high, only color sets with extremely high $\Delta S$ are above the threshold and thus there is very little variability in the values as is indicated by the blue curve. Thus, at very high $\Delta S$ criteria, there is too little variability in capacity for its correlations with total variation and entropy to be interpreted. Moreover, prior work has shown that people's accuracy at interpreting encoding mappings change very little for palettes with $\Delta S$ values higher than 0.7 \cite{schloss2021}. Due to these observations in addition to the lack of a principled way to pick a threshold for our data, we opted not to use threshold capacity.


In the case where $n$ (number of concepts in a set) is large, it is not feasible to compute max, mean, median, (or threshold) via exhaustive computation of all possible semantic distances. So, we considered two options:
(1) Compute semantic distance of the concept set with respect to the full feature library. In other words, extend the definition of generalized semantic distance from Section~\ref{sec:semdistthry} to this work. 
(2) Solve an assignment problem using the full feature library. Compute semantic distance of the resulting feature set. This is the \emph{max capacity} we ended up using.
Both of these approaches are scalable and computable even for large $n$. The problem with the first option is that as the feature library gets larger, we would want capacity to increase; having more features can make design easier. However, the opposite happens. As more features are added, generalized semantic distance gets smaller because perturbing the feature-concept associations leads to a large number of equally good assignments, which spreads out the probability distribution.
We used the second option (max capacity) because it only involves computing semantic distance of a feature set (size $n$) and will not decrease if we add more features to the feature library. 


\subsection{Total Variation}\label{sec:TV}
\begin{figure}[!ht]
 \centering
 \includegraphics[width=.8\columnwidth]{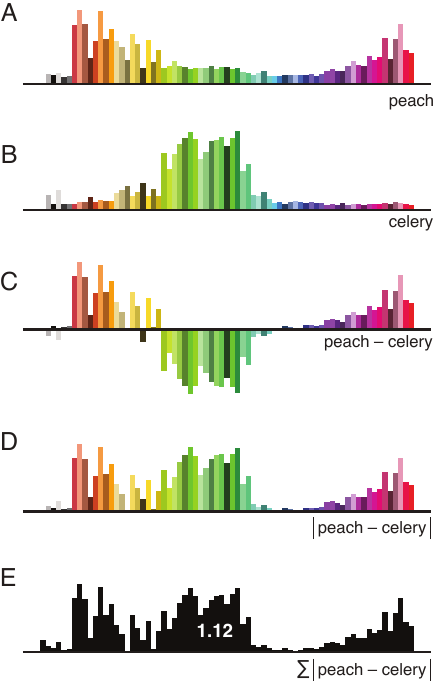}
 \caption{The color-concept distributions for peach (A) and celery (B), the difference (C) between those two distributions, the absolute value (D) of that difference, and the sum of the elements in the result (E). Half of the final number would be the TV computed for peach and celery.}
 \label{fig:peach_minus_celery}
\end{figure}

Computing total variation distance (TV) is computed by taking the difference between two color-concept association distributions (Figure \ref{fig:peach_minus_celery}C), taking the absolute value of that difference (Figure \ref{fig:peach_minus_celery}D), summing over the elements from the result (Figure \ref{fig:peach_minus_celery}E), and halving the resulting value (not shown). This is equivalent to taking half of the L1-norm of the difference between two color-concept association distributions.

\subsection{Generalized Total Variation} \label{sec:GTV}

Total variation distance (TV) has the following statistical interpretation. Suppose the true color distribution for a given concept is $p(\cdot)$. That is, for any color $i$, $p(i)$ is the normalized association rating between our given concept and color $i$. The true $p$ is unknown, but we have two candidate distributions $p_1$ and $p_2$ (each with prior probability $\frac{1}{2}$). Our task is to determine whether the true distribution is more likely to be $p_1$ or $p_2$, based on observing a single sample from $p$. In other words, the only information we have is a color $i$ drawn at random according to $p$. To minimize our probability of error, we should pick the \emph{likeliest} candidate, which is the one for which the probability of observing $i$ is largest. In other words, if $p_1(i) > p_2(i)$, we would infer that $p = p_1$, and vice versa. On average, the probability of estimating correctly using this maximum-likelihood estimator is given by:
\begin{align*}
\mathrm{prob}(\text{error}) &= 1-\frac{1}{2}\sum_i \max(p_1(i),p_2(i)) \\
&= 1 - \frac{1}{2}\sum_i \left( \frac{p_1(i)+p_2(i)}{2} + \left|\frac{p_1(i)-p_2(i)}{2} \right|\right) \\
&= 1 - \frac{1}{4}\sum_i p_1(i) - \frac{1}{4}\sum_i p_2(i) - \frac{1}{4} \sum_i \left| p_1(i) - p_2(i) \right| \\
&= \frac{1}{2} - \frac{1}{2}\text{TV}.
\end{align*}
When $\text{TV}=0$, the two distributions are identical, and $\text{prob}(\text{error}) = \frac{1}{2}$, so we are at chance. If $\text{TV}=1$, the maximum possible value of TV, $\text{prob}(\text{error}) = 0$. This corresponds to the case where the probability distributions have \emph{disjoint support}; for every $i$, either $p_1(i)=0$ or $p_2(i)=0$. So when $i$ is observed, we know with certainty which of the $p_j$ must have generated it.

If we generalize the statistical interpretation of TV above in a natural way, we obtain what we call \emph{generalized total variation} (GTV). Specifically, if we instead have $n$ candidate distributions $p_1,\dots,p_n$, the average probability of error when picking the likeliest distribution from a single sample is
\begin{align*}
\mathrm{prob}(\text{error}) &= 1-\frac{1}{n}\sum_i \max(p_1(i),\dots,p_n(i)) \\
&= \left(1 - \frac{1}{n}\right) - \frac{1}{n}\text{GTV}.
\end{align*}
In other words, we define GTV as:
\[
\text{GTV} = -1 + \sum_i \max(p_1(i),\dots,p_n(i))
\]
When $\text{GTV}=0$, again the distributions are all identical, and our probability of error is $1-\frac{1}{n}$ because we are at chance (we only have a $\frac{1}{n}$ chance of choosing correctly). When $\text{GTV}=n-1$, its maximum possible value, the $n$ distributions have mutually disjoint support, so for every $i$, at most one of the $p_j(i)$ is nonzero, which allows us to know with certainty which $p_j$ generated the sample, and the probability of error is zero.

When $n=2$, we get $\text{GTV}=\text{TV}$, which generalizes the standard definition of total variation distance. The trivial case $n=1$ also makes sense, because here $\text{GTV}=0$ and the probability of error is also zero.

Although there are many notions of \emph{statistical distance} for comparing pairs of distributions, TV being among the simplest, there are very few published techniques for comparing several distributions. One example is the Fr\'{e}chet mean~\cite[p.~136]{nielsen2012matrix}. To the best of our knowledge, our definition of generalized total variation is new.

\subsection{Distinctions between Semantic Distance, Semantic Contrast, and Accuracy in Experiment 2} \label{sec:SC_supp}
\textbf{Generalized semantic distance }is an operationalization of semantic discriminability with respect to an entire color palette. It estimates the probability of inferring the most-likely unique mapping between colors and concepts, compared with all other possible mappings in an encoding system (Section \ref{sec:semdistthry}). 

For example, imagine a visualization depicting three concepts---apple, banana, and blueberry---with three colors---red, yellow, and blue. If we were to solve an assignment problem using the balanced merit function and raw association scores between all three colors and all three concepts the most likely unique mapping would be: apple-red, banana-yellow, and blueberry-blue. If we repeatedly added random noise to the associations used as input and solved the assignment problem (using a Monte Carlo process), we can compute the probability of this most likely assignment compared to all other assignments (e.g., apple-yellow, banana-blue, blueberry-red, etc.). 

Generalized semantic distance estimates people's expectations about which colors map to which concepts, regardless of the ``correct'' assignment specified by the designer of an encoded system.  

\textbf{Semantic contrast} is also an operationalization of semantic distance, with respect to a single color relative to all other colors in a palette. It estimates the probability of inferring the optimal color-concept assignment for that color, where optimal is determined by an assignment problem using the balanced merit function. In the apple, banana, blueberry example above, the semantic contrast for, say, apple would be the proportion of times red was assigned to apple, compared to all other concepts. 

Like semantic distance, semantic contrast estimates which color maps to which concept, regardless of the ``correct'' assignment specified by the designer of an encoding system.

\textbf{Accuracy} is the degree to which people's interpretations of an coding system of a visualization match the encoded mapping---the ``correct'' assignment between colors and concepts specified by the designer. In some cases, the encoded mapping is explicitly specified by legends, labels, or other verbal descriptions, but encoded mappings can still exist in the absence of such explicit specifications. 

In Experiment 2 of the present study, we (designers) specified that the encoded mapping was the optimal mapping, generated by an assignment problem using the balanced merit function. Thus, accuracy for a given concept in a given concept set was the proportion of times participants chose the optimal color for that concept. 
We chose this definition of accuracy because previous work suggested assignments according to the balanced merit function match people's expectations (or inferred mappings). By matching people's expectations, these palettes should help people be as accurate as possible. We aimed to create conditions in which participants could be as accurate as possible to test the hypothesis that encoded mappings exist in which people can correctly interpret mappings for ``non-colorable'' concepts. 

The conditions of Experiment 2 present a special case in which semantic contrast can be thought of as a model for accuracy. This is because both semantic contrast and accuracy are defined with respect to assignments using the balanced merit function. However, this correspondence between accuracy and semantic contrast is not necessary. If the encoded mapping is defined in a different way that did not align with the optimal assignment (e.g., a default palette order from standard software), then semantic contrast would no longer estimate accuracy.
For example, one could define the encoded mapping in the earlier example as follows: apple-yellow, banana-blue, blueberry-red. In this case, the semantic contrasts remain unchanged however people's accuracies would be close to 0 for all colors (in the absence of a legend) because the encoded mapping heavily conflicts with people's inferred mappings. Thus, semantic contrast is always an operationalization of semantic discriminability, but not necessarily an operationalization of interpretability. 



If one wanted a general operationalization of interpretability, they could adapt the method used to compute semantic contrast so that instead of estimating the probability of the optimal solution, it estimates the probability of the correct solution, according to the encoded mapping in an encoding system. 

Additionally, if it were the case that there was a way of defining merit scores used in the assignment problem that estimated people's inferred mappings better than the balanced merit function, then our metrics including generalized semantic distance and semantic contrast could be defined with respect to this merit function to better capture human behavior.

  \begin{figure*}[!ht]
 \includegraphics[width=1\linewidth]{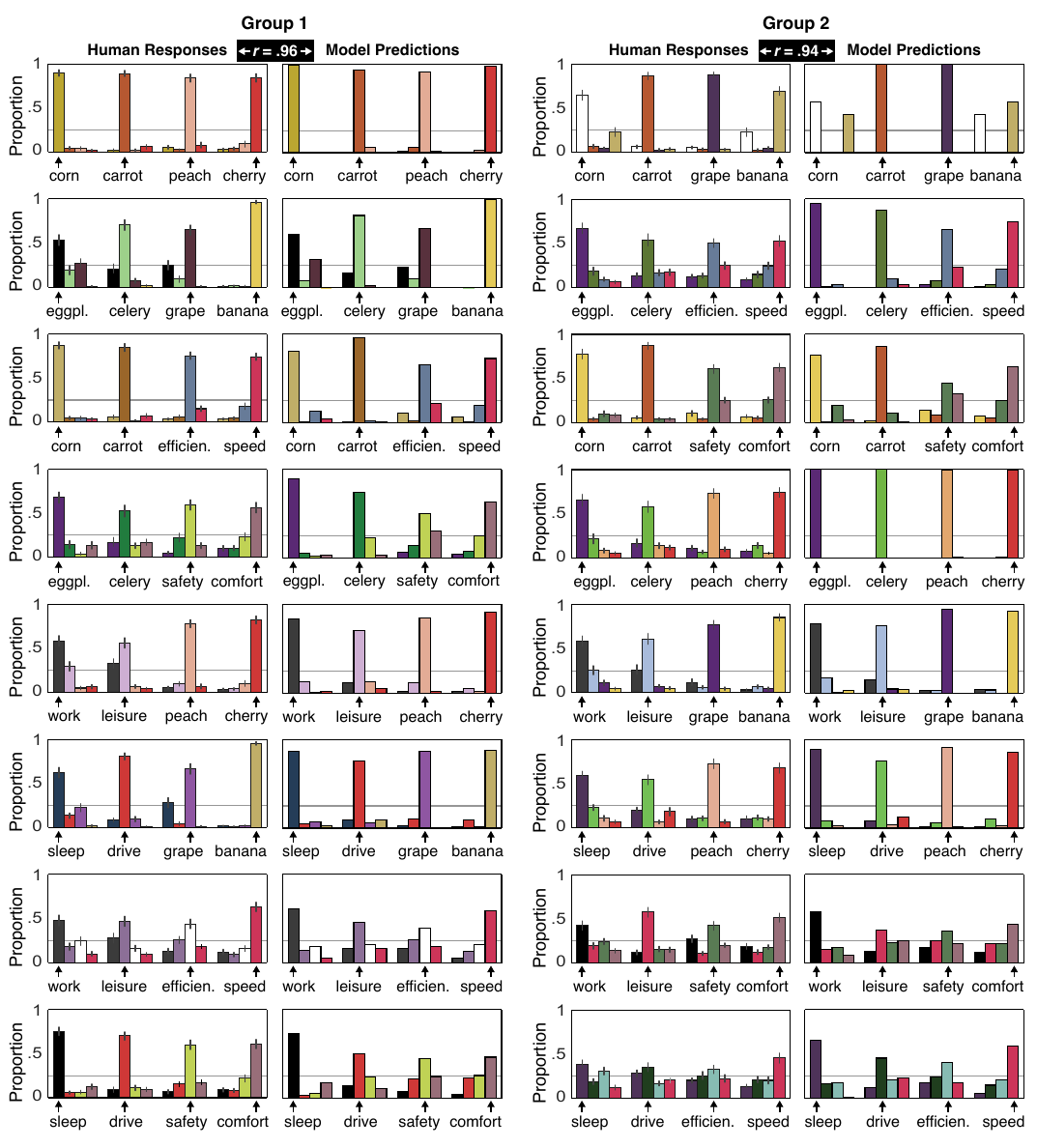}
 \caption{Responses from participants (left) and assignment predictions based on Monte Carlo sampling with perturbed color-concept association (right) for participant groups 1 and 2. Each pair of bar plots corresponds to one of the color-concept sets from Experiment 2. The solid black line across each graph corresponds to making an assignment by chance (.25). The correct response for each object is marked along the x-axis by an arrow. Error bars represent the ± standard errors of the means.}
 \label{fig:modelVdataALL}
\end{figure*}

\begin{table*}[!hb]
\sisetup{round-mode=places}
\centering
\caption{Coordinates for the University of Wisconsin 71 (UW-71) colors in CIE 1931 xyY space and CIELAB color space. The white point used to convert between CIE 1931 xyY and CIELAB space was CIE Illuminant D65 (x = 0.313, y = 0.329, Y = 100). The 'Sorted Position' column indicates the index of the color when the colors are sorted by hue angle and chroma as can be seen in Figure \ref{fig:assoc} and Figures \ref{fig:all_concepts}A and \ref{fig:all_concepts}B. }
\label{table:UW_71_colors}
\renewcommand{\arraystretch}{0.9}
\begin{tabular}
{c*{1}{
    S[round-precision=3]
    S[round-precision=3]
    S[round-precision=3]
    S[round-precision=2]
    S[round-precision=2]
    S[round-precision=2]
    S[round-precision=2]
}}
\toprule
\multicolumn{1}{c}{\textbf{Color}} &
\multicolumn{1}{c}{\textbf{Sorted Position}} &
\multicolumn{1}{c}{\textbf{x}} & \multicolumn{1}{c}{\textbf{y}} & \multicolumn{1}{c}{\textbf{Y}} & \multicolumn{1}{c}{\textbf{L*}} & \multicolumn{1}{c}{\textbf{a*}} & \multicolumn{1}{c}{\textbf{b*}}  \\
\midrule
1 & 50 & 0.17813 & 0.14021 & 18.419 & 50 & 28.891 & -73.589 \\
2 &53 & 0.1742 & 0.082514 & 4.4155 & 25 & 53.857 & -72.28 \\
3 & 54 & 0.21726 & 0.13588 & 18.419 & 50 & 53.857 & -72.28 \\
4 &55& 0.2591 & 0.13088 & 18.419 & 50 & 78.822 & -70.972\\
5 &46& 0.18715 & 0.19157 & 18.419 & 50 & 2.6168 & -49.931\\
6 &51& 0.19063 & 0.1298 & 4.4155 & 25 & 27.583 & -48.623\\
7&52 & 0.23145 & 0.18448 & 8.419 & 50 & 27.583 & -48.623\\
8&57 & 0.25495 & 0.12284 & 4.4155 & 25 & 52.548 & -47.315\\
9 &56 & 0.27872 & 0.17635 & 18.419 & 50 & 52.548 & -47.315\\
10&61  & 0.32783 & 0.1674 & 18.419 & 50 & 77.514 & -46.006\\
11&45  & 0.22397 & 0.28399 & 48.278 & 75 & -23.657 & -26.274\\
12&47  & 0.2081 & 0.21415 & 4.4155 & 25 & 1.3084 & -24.966\\
13&48  & 0.24471 & 0.25395 & 18.419 & 50 & 1.3084 & -24.966\\
14&49  & 0.26261 & 0.27354 & 48.278 & 75 & 1.3084 & -24.966\\
15&58  & 0.28644 & 0.19884  & 4.4155 & 25 & 26.274 & -23.657\\
16&59  & 0.29797 & 0.24051  & 18.419 & 50 & 26.274 & -23.657\\
17&60  & 0.30272 & 0.2622  & 48.278  & 75 & 26.274 & -23.657\\
18&62  & 0.36941 & 0.18108  & 4.4155 & 25  & 51.24 & -22.349\\
19&63  & 0.35288 & 0.2259 & 18.419 & 50 & 51.24 & -22.349\\
20&64  & 0.40795 & 0.21059 & 18.419 & 50 & 76.206 & -21.041\\
21&44  & 0.23784 & 0.35662 & 72.065 & 88 & -49.931 & -2.6168\\
22&43 & 0.25332 & 0.35108 & 18.419 & 50 & -24.966 & -1.3084\\
23&42 & 0.26938 & 0.34523 & 48.278 & 75 & -24.966 & -1.3084\\
24&41 & 0.27473 & 0.34327 & 72.065 & 88 & -24.966 & -1.3084\\
25&6 & 0.313 & 0.3290 & 0 & 0 & 0 & 0\\
26&5 & 0.31273 & 0.32902 & 4.4155 & 25 & 0 & 0\\
27&4 & 0.31273 & 0.32902 & 18.419 & 50 & 0 & 0 \\
28&3 & 0.31273 & 0.32902 & 48.278 & 75 & 0 & 0 \\
29&1 & 0.31273 & 0.32902 & 100.00 &  100 & 0 & 0 \\
30&2 & 0.31273 & 0.32902 & 72.065 & 88 & 0 & 0 \\
31&67 & 0.41044 & 0.2905 & 4.4155 & 25 & 24.966 & 1.3084 \\
32&68 & 0.37353 & 0.30534 & 18.419 & 50 & 24.966 & 1.3084\\
33&69 & 0.3568 & 0.31196 & 48.278 & 75 & 24.966 & 1.3084 \\
34&66 & 0.43376 & 0.28095 & 18.419 & 50 & 49.931 & 2.6168 \\
35&65 & 0.49181 & 0.25666  & 18.419 & 50 & 74.897 & 3.9252\\ 
36 &40& 0.27022 & 0.43268 & 48.278 & 75 & -51.24 & 22.349 \\
37&39 & 0.27623 & 0.41829 & 72.065 & 88 & -51.24 & 22.349 \\
38 &36& 0.3075 & 0.52435 & 4.4155 & 25  & -26.274 & 23.657\\
39 &33& 0.31561 & 0.44408 & 18.419 & 50  &  -26.274 & 23.657\\
40 &32& 0.31654 & 0.41004 & 48.278 & 75 & -26.274 & 23.657\\
41&31 & 0.31652 & 0.39917 & 72.065 & 88 & -26.274 & 23.657\\
42&20 & 0.41791 & 0.4496 & 4.4155 & 25 & -1.3084 & 24.966\\
43&17 & 0.38179 & 0.40728 & 18.419 & 50 & -1.3084 & 24.966\\
44&16 & 0.36355 & 0.38634 & 48.278 & 75 & -1.3084 & 24.966\\
45&15 & 0.35735 & 0.37928 & 72.065 & 88 & -1.3084 & 24.966\\
46 &10& 0.52174 & 0.37656 & 4.4155 & 25 & 23.657 & 26.274\\
47 &9& 0.44682 & 0.36993 & 18.419 & 50 & 23.657 & 26.274\\
48 &8& 0.41032 & 0.36214 & 48.278 & 75 & 23.657 & 26.274\\
49 &7& 0.50873 & 0.33341 & 18.419 & 50 & 48.623 & 27.583\\
50 &70 & 0.56618 & 0.29873 & 18.419 & 50 & 73.589 & 28.891\\
51&38 & 0.29736 & 0.57731 & 18.419 & 50 & -52.548 & 47.315\\
52&35 & 0.31049 & 0.50294 & 48.278 & 75 & -52.548 & 47.315\\
53 &34& 0.31313 & 0.47888 & 72.065 & 88 & -52.548 & 47.315\\
54 &28& 0.36753 & 0.52508 & 18.419 & 50 & -27.583 & 48.623\\
55 &28& 0.35998 & 0.47135 & 48.278 & 75 & -27.583 & 48.623\\
56&26 & 0.35593 & 0.45307 & 72.065 & 88 & -27.583 & 48.623\\
57 & 22& 0.43671 & 0.47238 & 18.419 & 50 & -2.6168 & 49.931\\
58 &19& 0.4092 & 0.43925 & 48.278 & 75 & -2.6168 & 49.931\\
59 &18& 0.39861 & 0.42682 & 72.065 & 88 & -2.6168 & 49.931\\
60&13 & 0.50246 & 0.42134 & 18.419 & 50 & 22.349 & 51.24\\
61&12 & 0.4572 & 0.40735 & 48.278 & 75 & 22.349 & 51.24\\
62 &11 & 0.56315 & 0.37345 & 18.419 & 50 & 47.315 & 52.548\\
63 &71& 0.61793 & 0.32963 & 18.419 & 50 & 72.28 & 53.857\\
64 &37& 0.30023 & 0.56426 & 72.065 & 88 & -78.822 & 70.972\\
65 &29& 0.34264 & 0.56147 & 48.278 & 75 & -53.857 & 72.28\\
66&30 & 0.34455 & 0.53221 & 72.065 & 88 & -53.857 & 72.28\\
67 &24& 0.39381 & 0.52123 & 48.278 & 75 & -28.891 & 73.589\\
68&25 & 0.38886 & 0.49964 & 72.065 & 88 & -28.891 & 73.589\\
69 &23& 0.44388 & 0.48131 & 48.278 & 75 & -3.9252 & 74.897\\
70&21 & 0.43244 & 0.46714 & 72.065 & 88 & -3.9252 & 74.897\\
71&14 & 0.49196 & 0.4425 & 48.278 & 75 & 21.041 & 76.206\\
\bottomrule
\end{tabular}
\end{table*}

\end{document}